\documentclass[fleqn,usenatbib]{mnras}
\usepackage{graphicx}
\usepackage{natbib}
\usepackage{epstopdf}
\usepackage{amssymb}
\usepackage{array, makecell}
\usepackage{amsmath}
\usepackage{mathptmx}
\usepackage{booktabs}
\usepackage{hyperref}
\usepackage{float}
\usepackage{subfig}

\newcommand{\msun}{\,{\rm M_{\odot}}}

\newcommand{\cm}{\,{\rm cm}}
\newcommand{\s}{\,{\rm s}}	
\newcommand{\erg}{\,{\rm erg}}

\newcommand{\Ha}{{\it H_{0.1}}}
\newcommand{\Hb}{{\it H_{1}}}
\newcommand{\Wa}{{\it W_{0.1}}}
\newcommand{\Wb}{{\it W_{1}}}
\newcommand{\Sa}{{\it S_{0.1}}}
\newcommand{\Sb}{{\it S_{1}}}
\newcommand{\Wc}{{\it W_{\infty}}}
\newcommand{\Sc}{{\it S_{\infty}}}
\newcommand{\sigl}{{\sigma_{0,l}}}
\newcommand{\sigh}{{\sigma_{0,h}}}
\newcommand{\sigz}{{<\sigma_0>}}

\usepackage[normalem]{ulem}
\usepackage{xcolor}
\definecolor{new_color}{rgb}{1.0, 0.43, 0.3}

\title[GRBs from intermittent mildly magnetized jets]{Intermittent mildly magnetized jets as the source of GRBs}
\author[O. Gottlieb et al.]{
	Ore Gottlieb\thanks{oregottlieb@mail.tau.ac.il},
	Omer Bromberg,
	Amir Levinson,
	Ehud Nakar
	\\
	{School of Physics and Astronomy, Tel Aviv University, Tel Aviv 69978, Israel}
}
\pubyear{2021}
\begin{document}
	\label{firstpage}
	\pagerange{\pageref{firstpage}--\pageref{lastpage}}
	\maketitle	
	\begin{abstract}

Gamma-ray bursts (GRBs) are powered by relativistic jets that exhibit intermittency over a broad range of timescales - from $ \sim $ ms to seconds.  Previous numerical studies have shown that hydrodynamic (i.e., unmagnetized) jets that are expelled from a variable engine are subject to strong mixing of jet and cocoon material, which strongly inhibits the GRB emission. In this paper we conduct 3D RMHD simulations of mildly magnetized jets with power modulation over durations of 0.1 s and 1 s, and a steady magnetic field at injection. We find that when the jet magnetization at the launching site is $\sigma \sim 0.1$, the initial magnetization is amplified by shocks formed in the flow to the point where it strongly suppresses baryon loading. We estimate that a significant contamination can be avoided if the magnetic energy at injection constitutes at least a few percent of the jet energy. The variability timescales of the jet after it breaks out of the star are then governed by the injection cycles rather than by the mixing process, suggesting that in practice jet injection should fluctuate on timescales as short as  $ \sim 10 $ ms in order to account for the observed light curves. Better stability is found for jets with shorter modulations. We conclude that for sufficiently hot jets, the Lorentz factor near the photosphere can be high enough to allow efficient photospheric emission. Our results imply that jets with $ 10^{-2} < \sigma < 1 $ injected by a variable engine with $ \sim 10 $ ms duty cycle are plausible sources of long GRBs.

	\end{abstract}
	
	\section{Introduction}
	\label{sec:introduction}
	
	Despite years of extensive research, the underlying mechanism of the prompt gamma-ray emission is still one of the most fundamental mysteries of Gamma-ray bursts (GRBs).
	The prompt emission is associated with several notable characteristics of GRBs,
	such as high radiation efficiency \citep[e.g.][]{Granot2006,Ioka2006,Zhang2007,Beniamini2016} and rapid temporal variability \citep[e.g.][]{Ramirez-Ruiz2000,Nakar2002b,Nakar2002c,MacLachlan2012,Bhat2013}. The latter could emerge from a variable engine \citep{Levinson1993,Sari1997,MacFadyen1998,Fenimore1999,Aloy2000}, or from the interplay between the jet and a high pressure cocoon which it inflates \citep[e.g.][]{Aloy2002,Matzner2003,Morsony2007,Gottlieb2019}. The jet-cocoon interplay induces the variability by virtue of forming local hydrodynamic instabilities along the jet-cocoon interface \citep[JCI;][]{Gottlieb2021} which render the jet structure irregular .
	The non-linear nature of the jet physics and the turbulent behavior of the cocoon imply that a complete study of the jet evolution to the emission zone can only be obtained through 3D simulations.
	
	A numerical and analytic work by \citet{Gottlieb2019} showed that continuously powered {\it hydrodynamic} (i.e., unmagnetized) jets yield highly efficient photospheric emission for any reasonable set of jet parameters. Their 3D simulations also showed that all hydrodynamic jets are subject to local hydrodynamic instabilities that grow along the collimation shock at the jet base \citep[see also][]{Meliani2010,Matsumoto2013a,Matsumoto2013,Matsumoto2019,Matsumoto2017,Toma2017,Gourgouliatos2018}. The growth of the instabilities leads to efficient mixing of jet and cocoon material along the JCI. The mixing is reflected in the light curve as variations in the radiation efficiency along the jet.
	Therefore, if the mixing is not too intense such that the terminal Lorentz factor is $ \gtrsim 100 $, as found in their simulations, the photospheric emission from {\it continuously} launched jets exhibit both high efficiency and rapid variability. \citet{Gottlieb2019} argued that an additional advantage of this model is that pair production in the downstream of the collimation shock serves as a thermostat, which leads to a spectral peak of the photospheric emission that is consistent with observations. This model however has two shortcomings. First, the degree of mixing changes with time, indicating that some temporal evolution is expected in the light curve, which has not been observed. Second, and more importantly, if there is no additional dissipation processes between the collimation shock and the photosphere, then the spectrum of the emission is expected to have an exponential cut-off above the spectral peak. This is inconsistent with the observed prompt emission broken power-law spectrum. A potential solution to this problem is an additional dissipation process that acts near the photosphere, such as internal shocks. \citet{Gottlieb2019} found that the variable mixing leads to internal shocks, however it is unclear whether those are strong enough to reshape the emerging spectrum.
	
	The naive expectation is that a variable engine will lead to strong efficient internal shocks. It is also reasonable to expect that the central engine operates intermittently as its dynamical time is on order of ms, much shorter than the burst duration. Thus, in a following study \cite{Gottlieb2020a} examined the effect of engine variability on the propagation of hydrodynamic jets and on the resulting emission. They carried out numerical simulations of  intermittently launched hydrodynamic jets where the jet power has high and low power episodes. Interestingly, they found that such jets are subject to very heavy baryonic entertainment from the cocoon, much heavier than the one seen in continuous jets. The heavy mixing emerges during the low power episodes as the high pressure of the cocoon ``squeezes" the region between the last high power episode and the next one, filling it with heavy cocoon material. As a result, the next powerful jet episode encounters the dense cocoon material and this interaction leads to the heavy mixing. Ultimately, the jet material launched during high-power episodes dissipates all its energy on pushing the dense cocoon material that stands in its way such that both components are heavily mixed during the process. Subsequently the loaded jet reaches its photosphere with a Lorentz factor that is too low to generate the prompt GRB emission. The conclusion of these two studies is that {\it continuous hydrodynamic} jets can potentially be the sources of GRBs, although it is unclear whether they can generate the observed spectrum, while {\it intermittent hydrodynamic} jets cannot generate GRBs.
	
	A different, and arguably more realistic, picture of GRB jets is that they are at least weakly magnetized, as the jet launching is likely to be driven by magnetic fields \citep{Blandford1977,Komissarov2001}.
	As magnetic fields are known for suppressing the growth of local hydrodynamic instabilities \citep[e.g.][]{Millas2017,Matsumoto2019}, magnetized jets would naturally yield different structures than those of hydrodynamic jets.
	\citet{Gottlieb2020b} performed 3D simulations of continuously launched weakly magnetized jets and found that the growth of instabilities is inhibited in jets with a magnetization of $ \sigma \gtrsim 10^{-2} $ \citep[see also][]{Matsumoto2020}. The suppression of the instabilities renders the jet more stable, thereby keeping its terminal Lorentz factor high to power an efficient photospheric emission, while avoiding the temporal evolution that results from the change in the mixing. However, such jets do not feature the temporal variability in their light curve, owing to the low baryon loading. To conclude, continuously injected magnetized jets generate efficient radiation however lack high variability, while intermittent hydrodynamic jets naturally provide persistent variability but lack the high radiative efficiency. 

	Intermittent jets with subdominant magnetic fields may resolve this issue by stabilizing the jets against the mixing, thereby allowing both efficient and highly variable emission.
	However, this is not necessarily the case as the dominant mixing process in continuous and intermittent jets is different. In continuous jets the mixing is induced by the growth of the instabilities along the JCI, and can be suppressed by sub-dominant magnetic fields. In intermittent hydrodynamic jets, most of the mixing originates in the interaction between the jet material at the front of the high-power episodes and the high pressure cocoon material that squeezes the low-power jet. This interaction is similar to the one that occurs at the head of the jet, where mixing is taking place in shock. This type of mixing is not necessarily affected by sub-dominant fields. Therefore, in order to test the effect of sub-dominant field on intermittent jets a full numerical analysis is required.
	In this work we perform 3D relativistic magnetohydrodynamic (RMHD) simulations of variable mildly magnetized ($ 10^{-2} \lesssim \sigma \lesssim 1 $) jets propagating in a dense medium, and study the evolution of such systems for the first time.
	We find that a substantial magnetic component enables intermittent jets to avoid heavy mixing while propagating inside the surrounding medium, and thus is likely to be essential in order to explain the GRB prompt emission.
	The outline of this paper is as follows. In \S\ref{sec:setup} we set up the numerical framework and present the models that we consider. In \S\ref{sec:stability} and \S\ref{sec:magnetization} we examine the stability and the magnetization of the jet, respectively. In \S\ref{sec:post} we discuss the expected post-breakout structure. We summarize and discuss the implications of our results in \S\ref{sec:discussion}.
	
	\section{Numerical Setup}\label{sec:setup}
	
	We perform a set of high resolution (see discussion and convergence tests in Appendix \ref{sec:app}) 3D simulations with \textsc{pluto} v4.2 \citep{Mignone2007}, using the RMHD module and a relativistic ideal gas equation of state.
	For the integration we employ a third order Runge-Kutta time stepping, piece-wise parabolic reconstruction and an HLL Riemann solver. To impose $ \nabla \cdot {\bf B} =0$ we make use of \textsc{pluto}'s constrained transport scheme.
	
	Our setup is based on models of continuously injected magnetic jets, $ {\it LM-2} $ and $ {\it LM-1} $ from \citet{Gottlieb2020b}, denoted here as models $ \Wc $ and $ \Sc $, respectively.
	In these models the jet is carrying a toroidal magnetic field and is injected with an initial Lorentz factor $ \Gamma_0 = 5 $ from a nozzle with a radius of $ 5\times 10^7 $ cm at an altitude of $ z_0 = 4 \times 10^8 $ cm into a non-rotating star with a mass $ M_\star = 10\msun $, a radius $ R_\star = 10^{11}\cm $ and a density profile:
	\begin{equation}\label{eq:star}
	\rho_\star(r) = \frac{2}{\pi}\times 10^{23}{\rm \frac{g}{cm}} r^{-2}\Bigg(\frac{R_\star-r}{R_\star}\Bigg)^3~.
	\end{equation}
	We define the toroidal magnetic field profile at the injection point following \citet{Mignone2009,Mignone2013,Gottlieb2020b}:
	\begin{equation}
	b_\phi = \sqrt{4\pi h_0\rho_j\sigma_0c^2}{\left\{\begin{array}{c} 2r/r_{j,0} \qquad\qquad\qquad\qquad\qquad 2r<r_{j,0} \\ \frac{r_{j,0}}{2r}\Big(1-\frac{(r-r_{j,0}/2)^2}{(r_{j,0}-r_{j,0}/2)^2}\Big) \qquad \frac{r_{j,0}}{2}<r<r_{j,0} \end{array}\right\}}~,
	\end{equation}
	where $ \rho_j $ is the jet's mass density and $\sigma_0$ is the peak magnetization at half of the nozzle radius.
	In models $ \Wc $ and $ \Sc $ the values of $ \sigma_0 $ are 0.01 and 0.1, respectively.
	The jet is injected hot with a specific enthalpy $ h_0 \equiv 1+4p_t/\rho_j c^2+b_\phi^2/4\pi\rho_j c^2 = 100(1+\sigma_0) $, where $ p_t $ is the thermal pressure. It expands conically to an opening angle of $ \theta_0 \approx 0.7/\Gamma_0 = 0.14 $ \citep{Mizuta2013,Harrison2018}, before it is collimated by the cocoon. The maximal terminal proper-velocity (if no mixing occurs) is thus $ \eta_0 \equiv \sqrt{\Gamma_0^2h_0^2-1} \approx 500(1+\sigma_0) $.
	
	To each model $ \Wc $ and $ \Sc $ we apply step-function modulations of 0.1 s and 1 s in the luminosity, such that it jumps between 20\% and 100\% of a total (two-sided) luminosity, $ L = 10^{50} \erg~\s^{-1} $.
	We stress that the central engine variability is likely to be on shorter timescales, but these are not feasible for 3D RMHD simulations over the range of length scales used in this work, given our available computational resources.
	To avoid strong currents on the lower boundary that result from similar jumps in the tangential magnetic field, we keep the injected magnetic field profile constant throughout the duration of the simulation, and vary only the injected $ \rho_j \propto L $. It implies that $\sigma_0\propto b_\phi^2 h_0^{-1} \rho_j^{-1} \propto L^{-1}$ in the low power episodes, $ \sigh $, is 5 times higher than in the high power episodes, $ \sigl $.
	We compare our results with jet models from our previous studies: (i) two continuously injected magnetic jets models, $ \Wc $ and $\Sc $ ($ {\it LM-2} $ and $ {\it LM-1} $ in \citealt{Gottlieb2020b}); (ii) two intermittent hydrodynamic models $ \Ha $ and $ \Hb $ (models $ {\it E} $ and $ {\it G} $ in \citealt{Gottlieb2020a}). The reference models maintain the same engine variability and hydrodynamic or magnetohydrodynamic (with $ \sigma_0 = \sigl $) parameters as those in our models.
	The rest of the parameters in our simulations are listed in Table \ref{table:models}.
	
	The numerical grid used in the simulations includes three patches on the $ \hat{x} $ and $ \hat{y} $ axes and one on the $ \hat{z} $-axis.
	The inner patch on the $ \hat{x} $ and $ \hat{y} $ axes is uniform with 640 cells in the inner $ |1.5\times 10^9|\cm $. The outer patches are logarithmic with 80 cells in each direction from $ |1.5\times 10^9|\cm $ up to $ |1.5\times 10^{10}|\cm $. On the $ \hat{z} $-direction we use 2000 uniform cells from $ z_{0} $ to $ 2R_\star=2\times10^{11} $ cm. In total we have $ 800\times 800\times 2000 = 1.28\times 10^9 $ cells. We provide convergence tests in Appendix \ref{sec:app}.
	
	\begin{table}
		\setlength{\tabcolsep}{10.5pt}
		\centering
		\begin{tabular}{ | l | c c c c c | }
			
			\hline
			Model & $ \sigl $ & $ \sigh $ & $ T $ [s] & $ t_b $ [s] & $ t_f $ [s] \\ \hline
			
			$ \Ha $ & 0 & 0 & 0.2 & 17 & 41 \\
			$ \Hb $ & 0 & 0 & 2.0 & 17 & 48 \\
			$ \Wc $ & $ 10^{-2} $ & $ 10^{-2} $ & $ \infty $ & 7 & 7 \\
			$ \Wa $ & $ 10^{-2} $ & $ 5\times 10^{-2} $ & 0.2 & 14 & 17.5 \\
			$ \Wb $ & $ 10^{-2} $ & $ 5\times 10^{-2} $ & 2.0 & 11 & 12 \\
			$ \Sc $ & $ 10^{-1} $ & $ 10^{-1} $ & $ \infty $ & 5 & 5 \\
			$ \Sa $ & $ 10^{-1} $ & $ 5\times 10^{-1} $ & 0.2 & 9 & 11 \\
			$ \Sb $ & $ 10^{-1} $ & $ 5\times 10^{-1} $ & 2.0 & 8 & 9 \\
			
			\hline
			
		\end{tabular}
		\hfill\break
		
		\caption{
			The models' parameters. $ \sigl $ and $ \sigh $ are the initial values of $ \sigma $ during the high and low power episodes, respectively. $ T $ is the time cycle, $ t_b $ is the breakout time of the forward shock from the star, and $ t_f $ is the time at which the simulation ends.
		}
		\label{table:models}
	\end{table}
	
	\section{Jet stability}
	\label{sec:stability}
	
	\begin{figure*}
		\centering
		\includegraphics[scale=0.22]{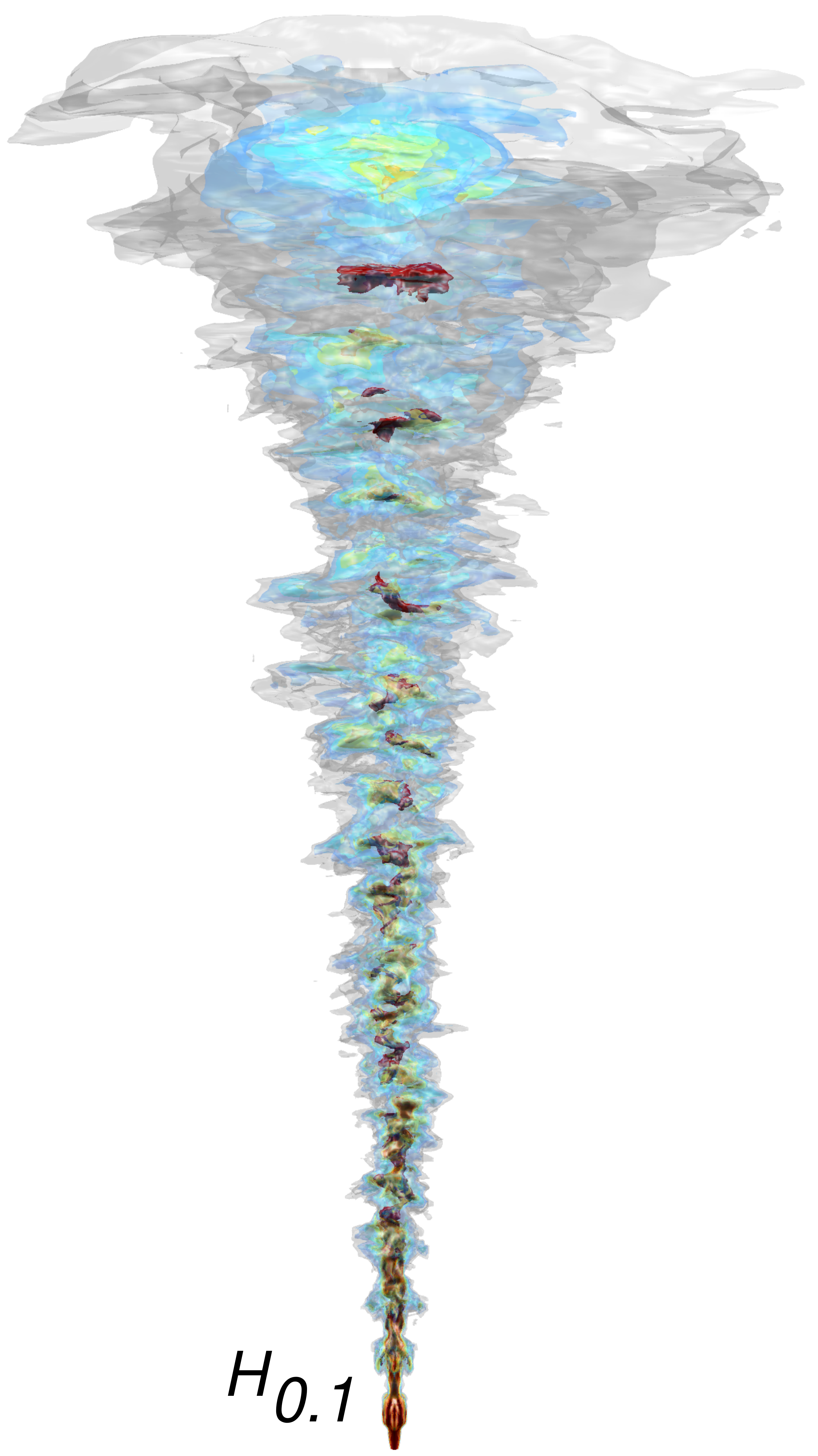}\hspace{12mm}
		\includegraphics[scale=0.22]{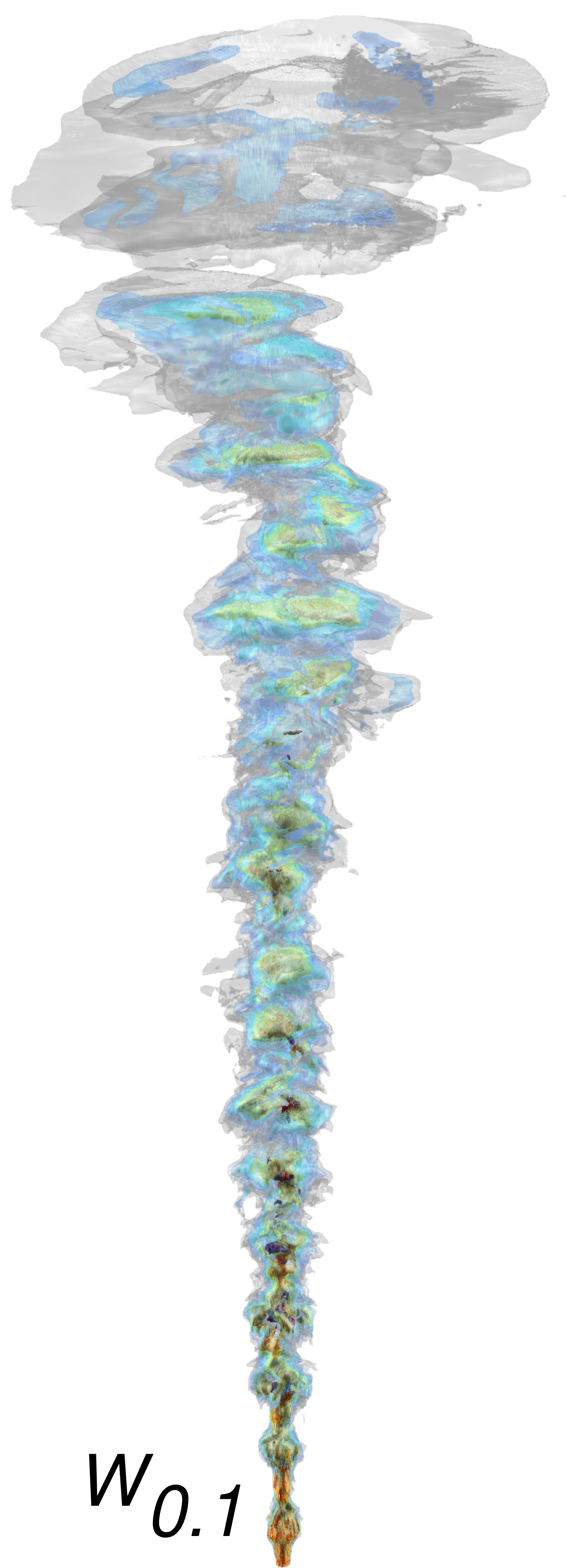}\hspace{10mm}
		\includegraphics[scale=0.22]{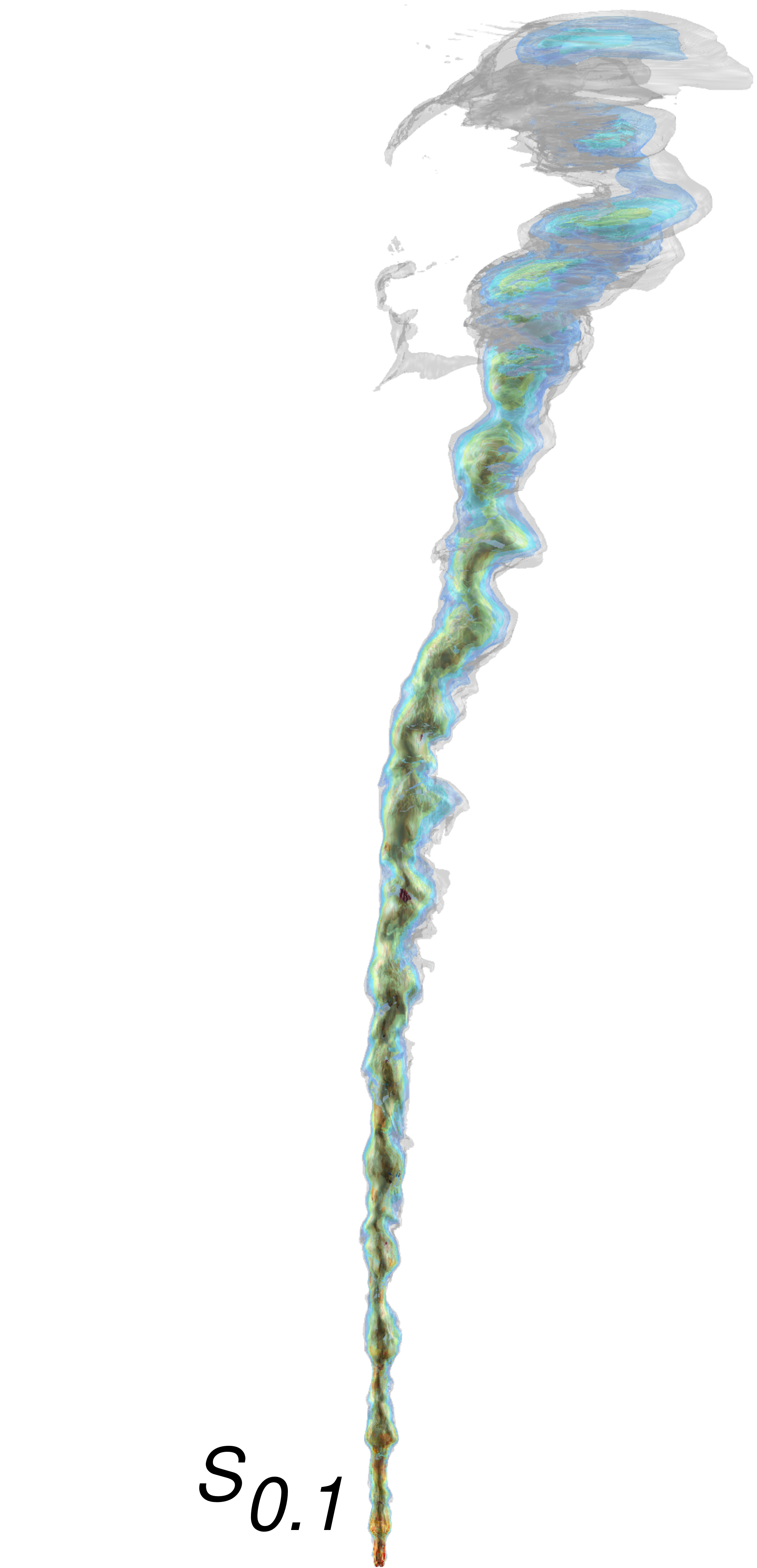}
		\includegraphics[scale=0.22,trim=3.65cm 0 0 0]{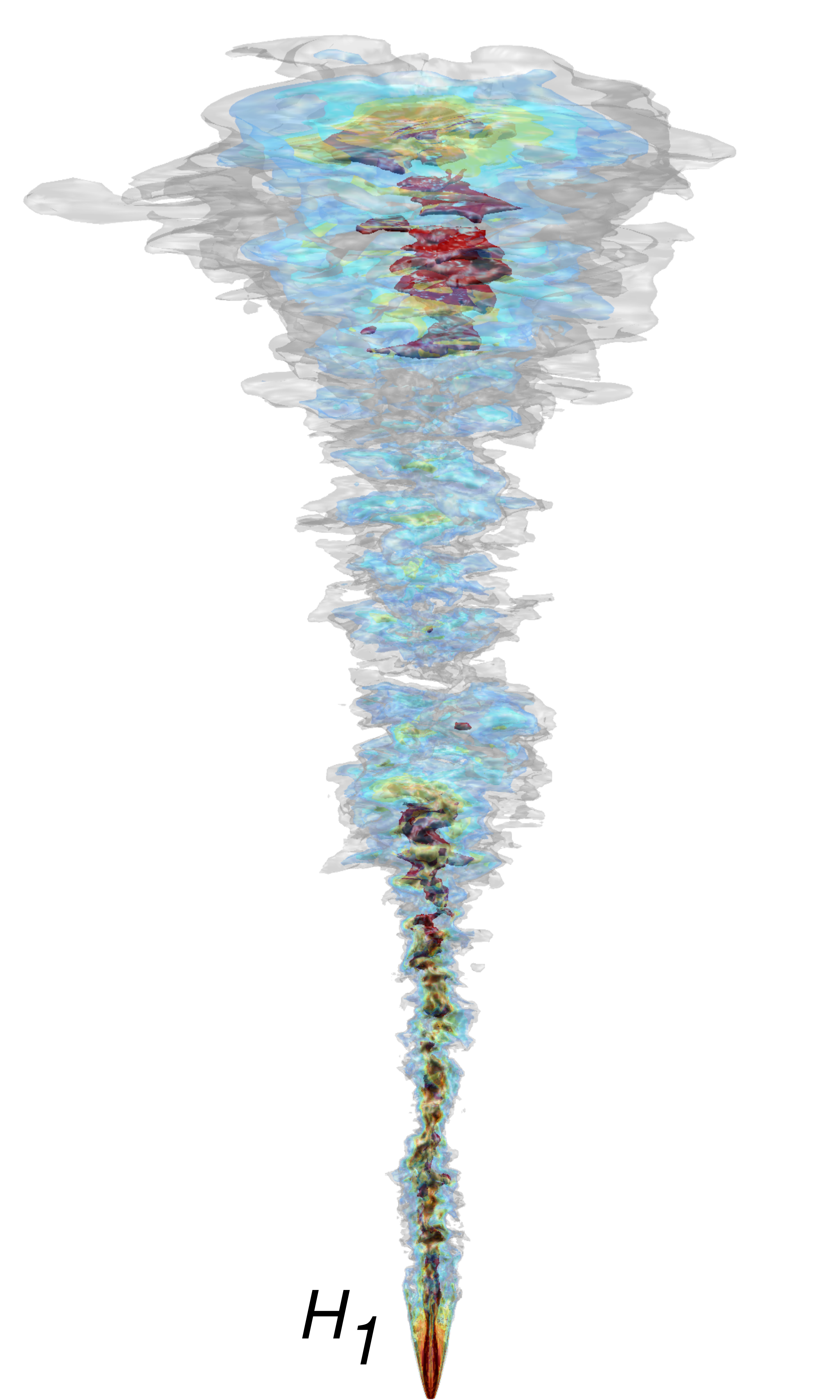}\hspace{15mm}
		\includegraphics[scale=0.22]{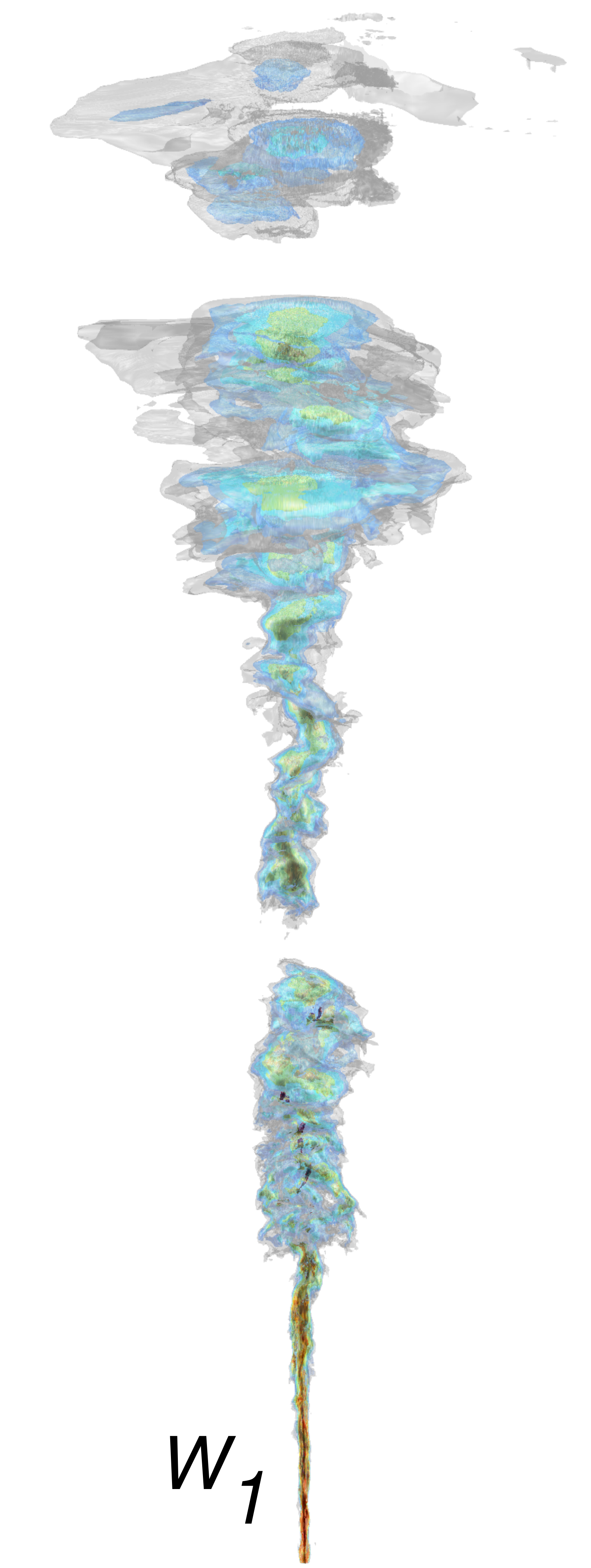}\hspace{17mm}
		\includegraphics[scale=0.22]{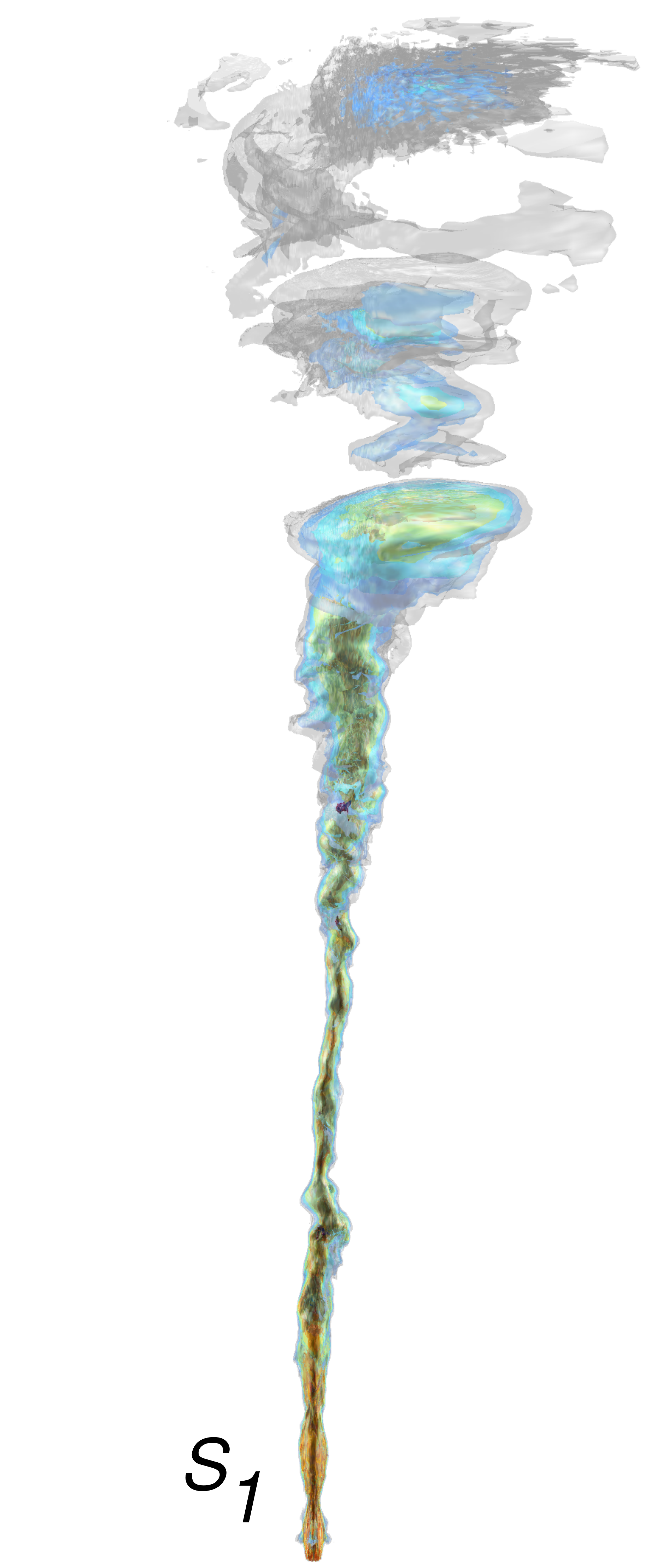}
		\caption[3D plots]{
			3D isocontours of $ {\rm log}(\eta_s) $ shortly after jet breakout from the star. Top: 0.1 s modulations $ \Ha $ (left), $ \Wa $ (middle) and $ \Sa $ (right). Bottom: 1 s modulations: $ \Hb $ (left), $ \Wb $ (middle) and $ \Sb $ (right).
			The values of $ {\rm log}(\eta_s) $ are 0.5 (gray), 0.8 (blue), 1.1 (green), 1.4 (yellow), 1.7 (orange) and 2 (red).
		}
		\label{fig:3d}
	\end{figure*}
	
	Continuous jet-cocoon interaction throughout the jet propagation in the star leads to mixing of jet and cocoon {material} along the JCI. The mixing reduces the terminal proper-velocity of the outflow to $ \eta_s \equiv \sqrt{\Gamma_s^2h_s^2-1} < \eta_0 $\footnote{The subscript $ s $ reflects the value of the quantity after mixing.} such that under intense mixing conditions, the jet becomes radiatively inefficient.
	Magnetic fields can stabilize the JCI and considerably reduce the mixing.
	Recent 3D RMHD simulations of continuously injected jets by \citet{Gottlieb2020b} have shown that the presence of a toroidal magnetic field can suppress the growth of the local hydrodynamic instabilities that emerge on the JCI. The magnetization degree required for jet stabilization depends on various jet parameters such as the jet power and initial opening angle. Low power or wider jets need stronger fields for stabilization whereas high power and narrower jets are more stable and thus the required magnetization is smaller. 
	Since in our models of modulated jets we vary the jet power and its magnetization such that $ L \propto \sigma_0^{-1} $, the stabilization of the jet interface may also vary between low and high power modes. We take the maximal luminosity (minimal $\sigma_0 $) to be that of the stable continuously injected jets in \citet{Gottlieb2020b}, so that the $ \sigma_0 $ values are at least as high as those which stabilize continuous jets. It is therefore expected that if the jet stability depends solely on the magnetic field's ability to suppress the growth of local hydrodynamic instabilities, our intermittent magnetized jets will also remain stable.
	If the mixing originates from the pulsation nature of the jet rather than the instabilities along the JCI, like in the case of intermittent hydrodynamic jets \citep{Gottlieb2020a}, then magnetic fields might have only a minor effect on the jet composition.

	At early times intermittent magnetic jets share similarities with both continuously launched magnetic jets and intermittent hydrodynamic jets.
	On one hand, they are supported by a toroidal magnetic field which stabilizes the jet and keeps $ \eta_s \approx \eta_0 $, similar to continuous jets \citep{Gottlieb2020b}.
	On the other hand, the modulations in the jet power destabilize the jet, since the low power episodes cannot support the jet against the confining cocoon, which collapses inwards towards the jet spine. All the energy of the previous high power episode has been used to accelerate heavy material with $ \eta_s \ll \eta_0 $ at the jet head, and the jet has to be rebuilt in the next powerful episode. When the front of the jet-cocoon system finally crosses a substantial part of the star, a structure of multiple mini-jets emerges, as was found in hydrodynamic jets \citep{Gottlieb2020a}.
    However, unlike hydrodynamic jets, here the jet magnetization which is kept at a level of $ \sigma \geq 10^{-2}$, has a stabilizing effect on both high power and low power episodes, and is able to prevent from the weaker jet episodes to completely mix with cocoon material.
	
	Figure \ref{fig:3d} shows iso-contours of $ {\rm log}(\eta_s) $ in intermittent jet models, shortly after the jet broke out from the star. Strong colors: red, orange and yellow mark higher $ \eta_s $ whereas blue-gray colors portray lower $ \eta_s $ values (see figure caption for accurate values).
	While all jets contain rather mixed material at their fronts, due to complete dissipation of the early modulations, the stability of the jet spine at lower parts differs between models.
	It is prominent that jets with strong fields (models $\Sa $ and $ \Sb $ on the right) are more stable than the rest of the models. The stability of the weaker magnetized jets (models $\Wa $ and $ \Wb $ in the middle) seems to be similar to that of the non-magnetized jets (models $\Ha $ and $ \Hb $ on the left). It is also shown that the magnetized jets exhibit some helical motion near their heads.
	In models $ \Sa $ and $ \Sb $ the jet head deviates from the symmetry axis by $ \sim 5^\circ$. This behavior can be attributed to two factors: (i) kink instability as these jets contain regions with $ \sigma \sim 1 $ \citep{Mizuno2009,Mizuno2012,Bromberg2016,Tchekhovskoy2016,Barniol2017}; (ii) asymmetric magnetic pressure that is built in the cocoon as we discuss next.
	
	\begin{figure}
		\centering
		\includegraphics[scale=0.4]{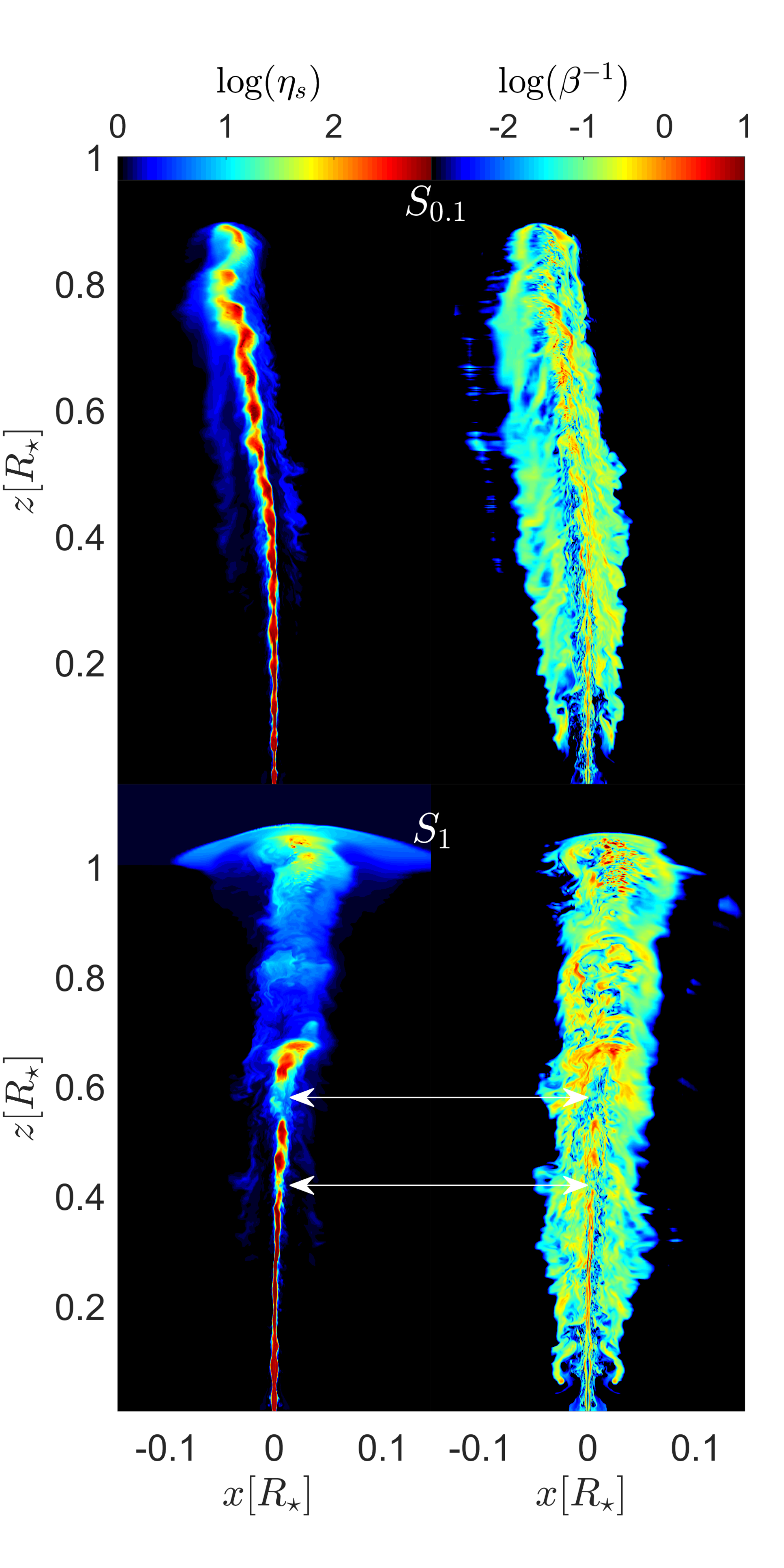}
		\caption[2D maps]{
			2D maps of $ \hat{x}-\hat{z} $ planes in models $ \Sa $ (top) and $ \Sb $ (bottom) upon jet breakout from the star. Shown are $ {\rm log}(\eta_s) $ (left) and magnetic to thermal pressure ratio $ {\rm log}(\beta^{-1}) $ (right). The white arrows demonstrate the relationship between low $ \eta_s $ regions and larger $ \beta $.
		}
		\label{fig:2d_maps}
	\end{figure}
	
	The stabilization of intermittent jets with $ \sigma \gtrsim 0.1 $ takes place when the jet crosses a significant part of the star. At that point the magnetic pressure in the jet prevents the heavy cocoon elements from fully mixing with the jet even in the low power sections, thereby keeping the jet spine rather uncontaminated by baryons.
	In Figure \ref{fig:2d_maps} we demonstrate how the magnetic pressure $ p_m $ dictates the jet stability and dynamics. It depicts meridian cuts of ${\rm log}(\eta_s)$ (left) and the reciprocal of the plasma beta: ${\rm log}(\beta^{-1})\equiv{\rm log}(p_m/p_t)$ (right) in models $ \Sa $ (top) and $ \Sb $ (bottom). One can see that jet elements with ultra-relativistic $ \eta_s \gtrsim 100 $ have a substantial $ p_m $ component ($ \beta \lesssim 1 $), implying that the magnetic pressure in the jet plays a key role in the stabilization of the low power episodes. In regions where $ p_m $ is subdominant, baryon entrainment from the cocoon penetrates into the jet and increases the mixing (white double arrows).
	In lower $ \sigma $ models $ p_m $ is dynamically subdominant throughout the jet, allowing strong mixing at all times.
	While the magnetic pressure in the jet suppresses baryon entrainment into the jet, the magnetic pressure in the JCI dictates the jet dynamics on large-scales. It can be seen that the magnetic pressure support is asymmetric between the two sides of the jet. As a result the jet tilts towards the side where $ \beta $ is higher, with larger asymmetry ($ \Sa $) leads to a larger deviation from the axis. The magnetic pressure in the JCI may also contribute to the jet head helical motion together with the expected kink instabilities.
	
	We stress that the value of $ \beta $ upon injection is not held constant between low and high power episodes. This comes as a result of holding the magnetic pressure fixed throughout the injection and varying the gas pressure as $ p_{t,0} \propto \rho_j \propto L $ in order to keep $ h_0 $ constant at all times. It then follows that the jet is launched with $ \beta \propto L $, so that high power episodes are less supported by the magnetic pressure upon injection. We find however that during the jet propagation in the star, $ \beta $ changes and becomes lower in the high power episodes than that in the low power ones. The reason is twofold: (i) High power episodes sustain stronger collimation shocks which amplify the magnetization in the jet (see \S\ref{sec:magnetization}), thereby decreasing the value of $ \beta $; (ii) High power jets are more stable \citep{Gottlieb2021} such that the high power episodes keep their $ \sigma_0 $ better than low power episodes in which $ \beta $ increases faster due to the mixing.
	
	Finally, when the modulations are longer ($ \Sb $), there are larger regions in the jet with low $ p_m $, allowing substantial baryon entrainment from the cocoon into the jet. In model $ \Sa $ only small parts along the jet have subdominant $ p_m $, and thus the mixing remains lower when the modulations are shorter.
	This is in contrast to intermittent hydrodynamic jets, in which longer modulations increase the jet stability \citep{Gottlieb2020a}. This behavior is due to the fact that hydrodynamic jets do not have any stabilization effect between high power episodes. Thus, small mini-jets dissipate their energy fast whereas larger ones can keep their structure for longer times. This lookout might be important as the time cycles in nature are more likely to be on dynamical timescales of an order of $ \sim $ ms, so that magnetic stabilization effect should be even more prominent.
	
	\section{Magnetization of the jet-cocoon system}
	\label{sec:magnetization}
	
	The intermittent jets are launched with toroidal magnetic fields into unmagnetized medium. As they propagate, they form cocoons of shocked magnetized jet and unmagnetized medium material, mixed together due to turbulence that grow in the hot plasma. The turbulence also generate poloidal fields, which remain subdominant in all regions where $ \sigma $ is high.
	
\begin{figure}
		\centering
		\includegraphics[scale=0.23]{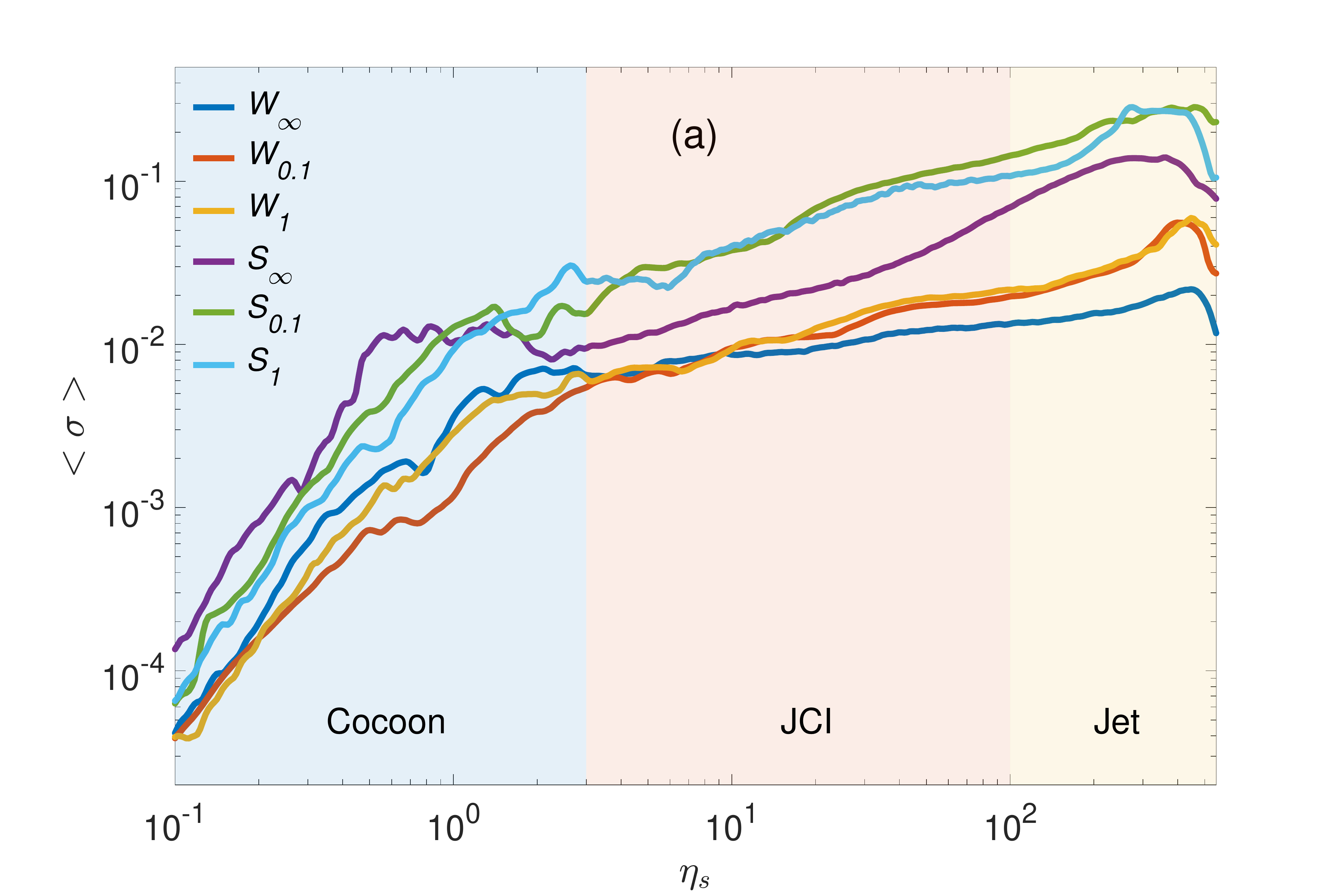}
		\includegraphics[scale=0.23]{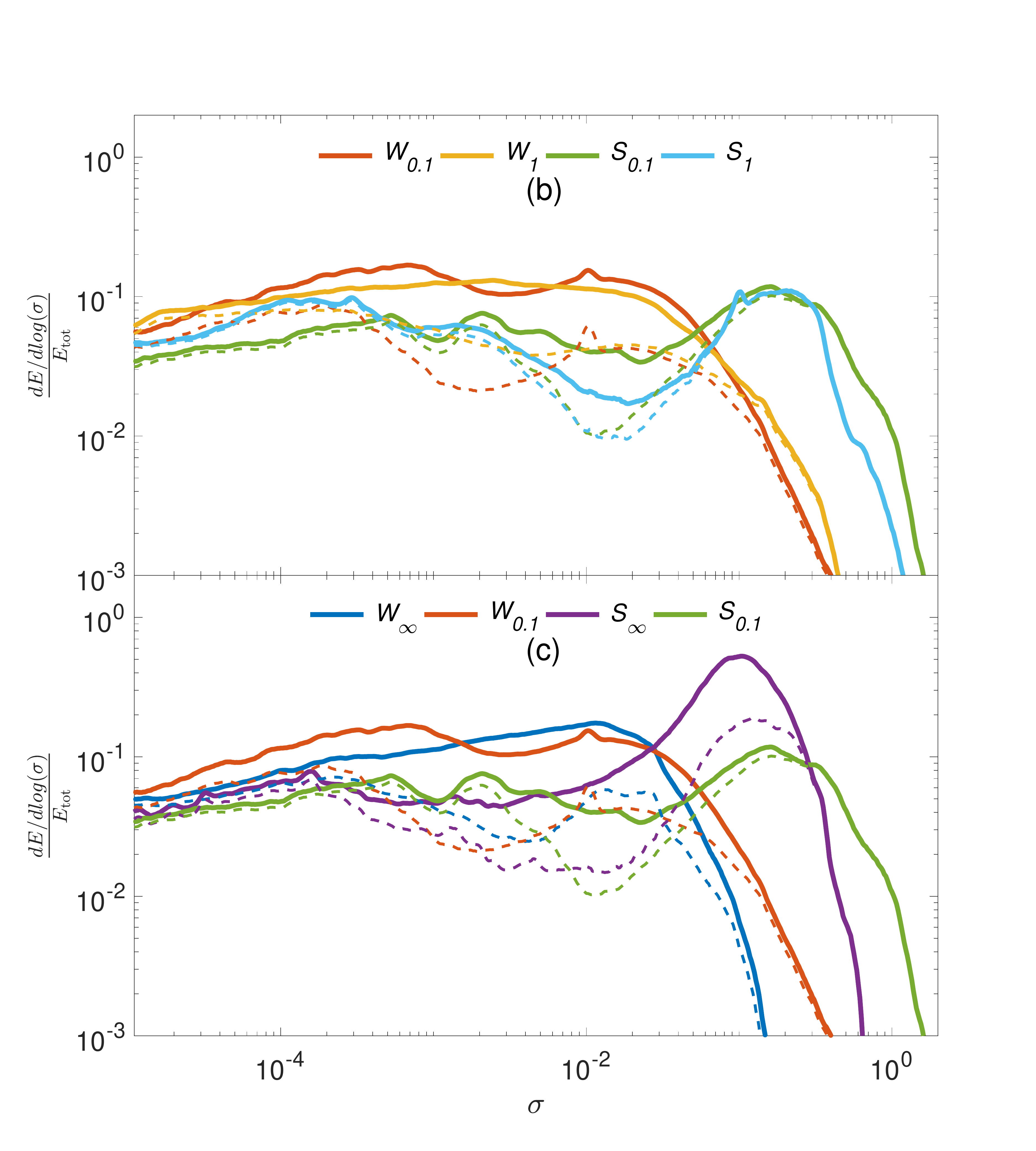}
		\caption[Energy as a function of $ \sigma $]{
			Panel (a) depicts the total energy (excluding rest-mass) weighted average of $ \sigma $ at each terminal proper-velocity value $ \eta_s $. Background colors mark the regions of the cocoon (light blue), JCI (light red) and jet (light yellow), as found by \citet{Gottlieb2021}.
			Panels (b) and (c) depict energy distributions as functions of $ \sigma $. Solid lines delineate the total distribution whereas dashed lines represent the distribution at $ z < \frac{1}{2}R_\star $.
			We present a comparison of different time cycles (b) and between continuous and intermittent jets (c).
		}
		\label{fig:sigma_dist}
	\end{figure}

	Figure \ref{fig:sigma_dist}a shows the average of $ \sigma $ weighted by the energy $ E $ (excluding rest-mass): $ <\sigma> = E^{-1} \int \sigma dE $, as a function of the terminal proper-velocity $ \eta_s $, taken at the last snapshot of each simulation. Regions in the plot that are associated with the jet, the cocoon or the JCI are highlighted with different background colors according to their $ \eta_s $.
	All models exhibit the same behavior, independent of the modulation time, featuring high $ \sigma $ in unmixed jet material (yellow background) and low $ \sigma $ in mixed cocoon material (blue background), in agreement with \S \ref{sec:stability}. The magnetization in the cocoon ($ \eta_s \lesssim 3 $) roughly scales as $ <\sigma> \sim \eta_s^2 $, demonstrating that shocked jet material (high $ \eta_s $) has substantially higher $ <\sigma> $ than the shocked stellar material (low $ \eta_s $). The dependency of $ <\sigma> $ on $ \eta_s $ becomes weaker at the JCI and the jet, until it peaks with a value $ \sim \sigz \equiv (\sigl+\sigh)/2 $ for the unmixed jet material at $ \eta_0 $.
	
	Figures \ref{fig:sigma_dist}b,c depict the normalized energy distribution per logarithmic scale of $ \sigma $ at the last snapshot of each simulation. Solid lines mark the total distribution in the system whereas dashed lines consider only material at the inner part of the star, $ z < \frac{1}{2}R_\star $.
	We find that qualitatively the total energy (solid lines) is distributed equally in logarithmic scales of $ \sigma $ up to $ \sigz $, the average of the injected magnetization peaks. The only exception is model $ \Sc $ which exhibits a bump at $ \sim \sigma_0 $, indicating that the jet remains stable and the mixing is low.
	More subtle differences are notable between models $ \Sa $, $ \Wa $ and $ \Wc $, with continuously injected ($ \Wc $) and strongly magnetized ($\Sa $) jets exhibit less mixing and lower cocoon/jet energy ratio.
	
	The left side of the distribution ($ \sigma \lesssim 10^{-3} $) marks the magnetization in the cocoon. Magnetized cocoon material flows into the cocoon from the jet head with initial magnetization of $ \sigma \sim \sigz $.
	Over time this material undergoes mixing with shocked medium material and its $\sigma$ decreases. Close to the jet base the magnetization is lower than $ \sigma \sim 10^{-5} $ and therefore the dashed lines and the solid lines coincide at this region.
	When the jet is intermittent, further mixing takes place as different jet episodes interact with each other, rendering the cocoon more energetic compared to the jet, as can be seen in Figure \ref{fig:sigma_dist}c.
	
	The right side of the distribution depicts the magnetization in the jet.
	The strong collimation shocks at the base of the jet amplify the magnetic field after the shock up to $ \sigma \sim 0.1 $.
	In models $ \Wa, \Wb $ and $ \Wc $ the amplification of the magnetic field is reflected by the extension of the energy distribution up to $ \sigma \gtrsim 0.1 > \sigh $, with all elements with $ \sigma \gtrsim 0.1 $ are located at $ z < \frac{1}{2}R_\star $ where the collimation shock resides, so that dashed and solid lines coincide.
	In models $ \Sa, \Sb $ and $ \Sc $ the amplified field is comparable with $ \sigma_0 $ and thus is not as prominent as in the lower $ \sigma $ jets. In model $ \Sc $ the dashed and solid lines do not coincide at $ \sigma \sim \sigma_0 $ since the mixing is minimal such that the magnetization remains high at all radii.
    When the central engine is variable, the amplification of the magnetic field also occurs in the internal shocks that are induced by the modulations, and are shown as longer high $ \sigma $ tails of the distributions of modulated jets.

	\section{Post-breakout evolution \& emission}
	\label{sec:post}
	
	\begin{figure*}
		\centering
		\includegraphics[scale=0.4]{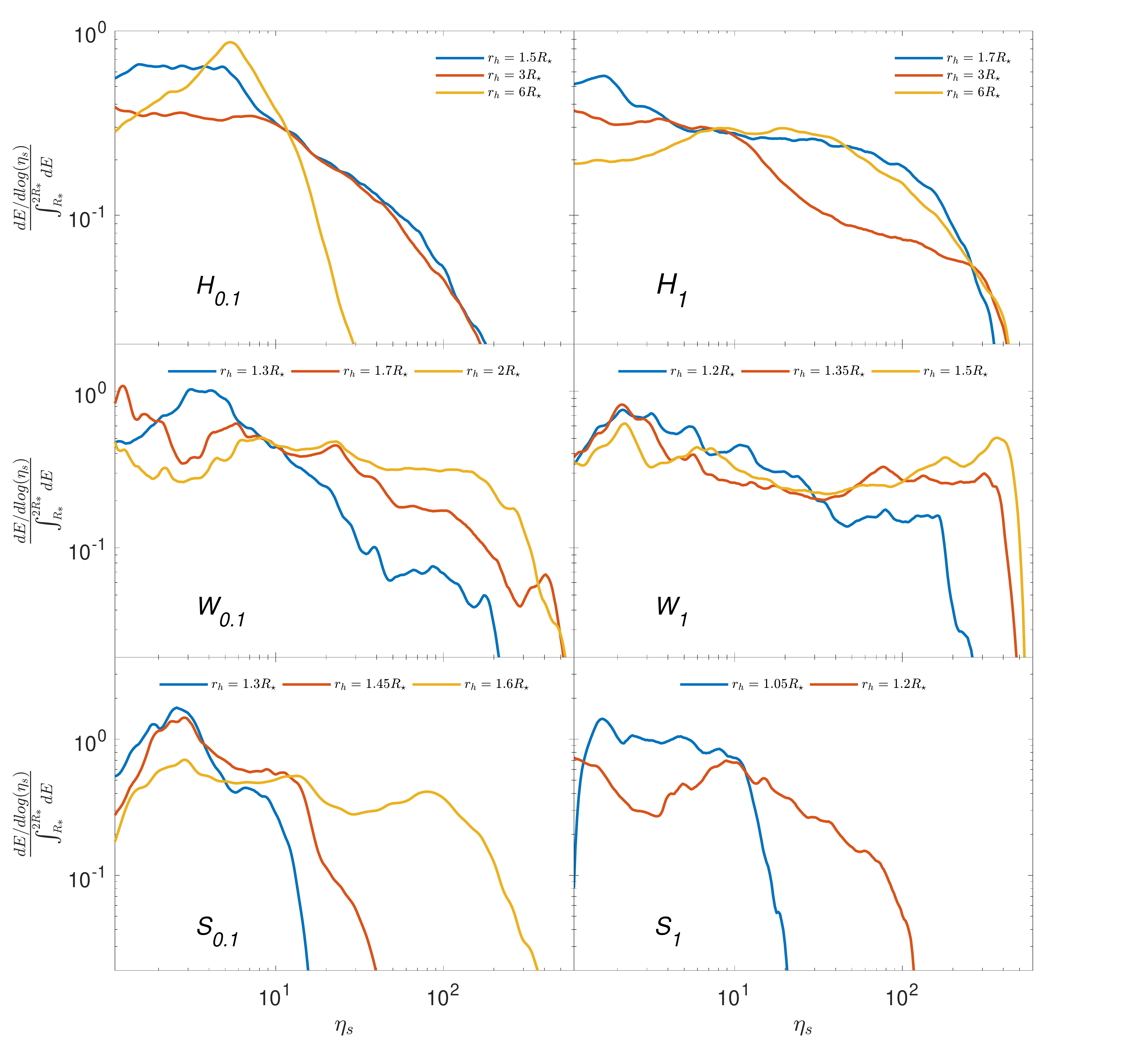}
		\caption[Distributions]{
			Energy distribution of matter outside the star as a function of $ \eta_s $. Shown are intermittent hydrodynamic models (top) and magnetized jets (center and bottom), with modulations of 0.1s (left) and 1s (right). Each model is shown at different times, manifested by the jet head location $ r_h $.
		}
		\label{fig:distributions}
	\end{figure*}
	
	The observed radiation is released at large radii, long after the jet breaks out of the star. Thus, in order to estimate the resulting emission one needs to follow the post-breakout jet evolution
	up to the point where the flow becomes homologous. However, tracking the post-breakout evolution of magnetized jets requires keeping very high resolution grid outside of the star. With our computational resources we can follow the evolution only to the point where the jet head reaches $ \sim 2R_\star $. In some simulations even our highest resolution turns out to be insufficient, and the numerical integration crashes soon after the jet breakout, limiting our ability to analyze the post-breakout behavior any further. Hence, in order to estimate the effect of the jet structure of intermittent magnetized jet on the observed emission, we compare the outflow structure at this limited range with the most stable intermittent hydrodynamic models $ \Ha $ and $ \Hb $. These simulations of hydrodynamic jets reach larger radii than magnetized jets. Thus we can use the comparison between magnetized and unmagnetized jets at early time together with comparison to the distribution of hydrodynamic jets at late times to learn about the expected late time evolution of the magnetized jets.
	
	Figure \ref{fig:distributions} depicts the energy distribution of material that broke out of the star as a function of $ \eta_s $, when the jet head is at different radii, $ r_h $. The figure compares the distributions of various hydrodynamic and magnetized intermittent jets.
	The intermittent hydrodynamic models contain predominant mildly-relativistic material at early times. The temporal evolution of the energy distribution of model $ \Ha $ shows two trends: (i) The distribution converges to a prominent peak, owing to internal shocks between the fast elements with the mildly-relativistic quasi-spherical cocoon in the front; (ii) The peak is shifted to higher velocities (blue to yellow curves), due to acceleration of the outflow by freshly less mixed elements that exit the star. However, even after $ \sim 30 $s of engine activity (corresponds to the line of $ r_h = 6R_\star $), the peak is at $ \eta_s \sim 10 $, implying that this model cannot generate a GRB.
	In model $ \Hb $ those trends are not observed due to a longer time cycle of the modulations which leads to longer times between internal shocks. The late time analysis conducted by \citet{Gottlieb2019} showed that a prominent peak of $ \eta_s \approx 30 $ is obtained when $ r_h $ is at a few dozen stellar radii (still below the photosphere), implying that jets with longer modulations cannot produce an efficient emission either.
	
	By contrast to the hydrodynamic models, $ \eta_s $ of intermittent magnetized jets recovers over time and converges to a rather flat energy distribution up to $ \eta_0 $. This comes as a result of a reduction in the mixing of matter that breaks out from the star at later times, as was also found in the post-breakout evolution of the continuously injected magnetized jets $ \Wc $ and $ \Sc $ \citep{Gottlieb2020b}.
	Therefore, while e.g. the distributions of $ \Hb $ and $ \Wa $ are rather similar when the jet breaks out (as also seen in Figure \ref{fig:3d}), the magnetized jet evolution shows an extension to higher $ \eta_s $ whereas the hydrodynamic jet converges to a peak at low $ \eta_s $.
	Furthermore, model $ \Wb $ exhibits a flat distribution of the outflow all the way to $ \eta_0 $ already when $ r_h = 1.5R_\star $.
	As we discussed in \S\ref{sec:stability}, the jets with the stronger magnetic fields (models $ \Sa $ and $ \Sb $) are the most stable ones.
	This is also demonstrated here with model $ \Sa $ being more stable than model $ \Wa $.
	However, the higher magnetization also causes more numerical noise, limiting our ability to analyze model $ \Sb $ far from the star, as its simulation crashes soon after the jet breakout.
	In summary, Figure \ref{fig:distributions} implies that the resulting photospheric emission from intermittent hydrodynamic jets will be inefficient, owing to low terminal $ \eta_s $, whereas magnetized jets become more stable over time, such that most of their energy lies in ultra-relativistic velocities to power an efficient photospheric emission.
	
	\section{Conclusions \& Discussion}
	\label{sec:discussion}
	
	Previous numerical studies have shown that hydrodynamic jets are subject to hydrodynamic instabilities that grow on the JCI boundary, induce mixing between the jet and the cocoon and give rise to rapid variability \citep{Gottlieb2019}. When magnetic fields are introduced in continuous flows, the growth of the instabilities is suppressed and thus these jets cannot account for the observed high variability of GRB light curves \citep{Gottlieb2020b}.
	Continuous jet injection over the crossing time of the stellar envelope seems unlikely given the dynamical time (ms) of the putative engine.
	Recent simulations by \citet{Gottlieb2020a} indicate that hydrodynamic jets with modulated power are subject to another type of mixing. In those jets the mixing emerges at the fronts of the high-power jet episodes as they slam into the heavy cocoon material that squeezes the low-power jet to fill up the regions between the high-power episodes. This mixing takes place via shocks in a manner that is similar to the mixing that arises at the jet head, as if each high-power episode develops its own head. We denote this type of mixing as ``head-like".
	This type of mixing gives rise to heavy loading that inhibits emission nearly completely, such that intermittent hydrodynamic jets are also inconsistent with GRB observations.
	Nonetheless, under the hypothesis that jets are produced by magnetic extraction of the BH (or magnetar) spin energy, the jet is likely to be 
	magnetized well inside the star, and the question remains as to how this might affect the mixing of modulated jets.

	In this paper we report on high resolution 3D simulations of the propagation of modulated, mildly magnetized ($0.01\le\sigma\le 0.5$) RMHD jets in a star.  The jets are injected with a steady, toroidal magnetic field and power modulations with duty cycles of 0.1 and 1 seconds (in two different experiments).
	We find that under these conditions, collimation and internal shocks in the jet may amplify the jet and cocoon magnetization to $ \sigma \gtrsim 0.1 $ and introduce mixing at the head, similar to intermittent hydrodynamic jets. However, the head-like mixing decreases substantially when the magnetization becomes large enough. The key physical property that reduces the mixing is the increased magnetic pressure during the low-power episodes that prevents the cocoon from filling the entire region between the high-power episodes.
	We conclude that if the magnetic energy constitutes at least a few percent of the jet energy, the magnetic pressure in the jet keeps the baryon load well below the level found in pure hydrodynamic jets, even during the low power episodes\footnote{Except for the front part of the outflow (of size $ \sim R_\star $) that broke out first.}.

    Due to numerical limitations we were unable to simulate the jet evolution outside of the star beyond $\sim 2$ stellar radii. For this reason we cannot study the late-time jet structure and its temporal evolution directly from the simulations, or provide a quantitative calculation of the resulting light curves as been done in \citet{Gottlieb2019,Gottlieb2020a}. A qualitative discussion of the above follows.

	In \S\ref{sec:post} we follow the jets to $ \sim 2R_\star $ and show that over time the energy distribution in $ {\rm log}(\eta_s) $ space converges to a rather flat profile, similar to the profiles of continuous hydrodynamic jets and flatter than profiles of intermittent hydrodynamic jets.
	Following \citet{Gottlieb2019} we calculate the average $ \eta_s $ value on the jet axis in models $ \Wa $ and $ \Wb $ above the collimation shock in the last snapshot of the simulation. 
	We find average values of $<\eta_s>(\Wa) \approx 270 $ and  $<\eta_s>(\Wb) \approx 220 $, which imply that both jets contain regions of ultra-relativistic terminal velocities in consistence with GRB data. The jet of model $\Wa$ with shorter modulations  has a typical higher $ \eta_s $ value, which indicates a higher stability.
	The measured $<\eta_s>$ values are larger by about an order of magnitude from the values measured in intermittent hydrodynamic jets \citep{Gottlieb2020b}. This implies that intermittent unmagnetized jets are less stable and have higher baryon contamination than magnetized jets.
    A similar analysis of higher $ \sigma $ jets cannot be preformed due to their deviation from the jet axis. The source of the shift, which is caused by an imbalance in the magnetic pressure at the cocoon is still unclear. However, since the jets in models $ \Sa $ and $ \Sb $ seem to be more stable than those in models $ \Wa $ and $ \Wb $, based on our analysis in \S\ref{sec:stability} and \S\ref{sec:post}, 
    we expect their $ <\eta_s>$ to be larger than 200 as well.
    
    The high $ \eta_s$ values and flat energy distributions in ${\rm log}(\eta_s)$ space, seen in intermittent mildly magnetized and continuous hydrodynamic jets, imply that the jets should share some similarities in their observable characteristics. \citet{Gottlieb2019} found that in their simulated steady hydrodynamic jets the photospheric radius and radiative efficiency were $ r_{\rm ph} \approx 10^{12} $ cm and $ \epsilon \gtrsim 0.5 $, respectively.
	Since in our simulations $ \sigma < 1 $, the specific enthalpy at the photosphere is likely to be dominated by the radiation thermal pressure, and thus intermittent magnetized jets are expected to show similarly high radiative efficiencies at their photosphere.
	
	An important difference between steady hydrodynamic outflows and intermittent mildly magnetized jets is the characteristic variability, which may affect the resulting emission via e.g., internal shocks. In continuous hydrodynamic models the variability is set by the physics of the instabilities that grow on the jet boundary and mix heavy material into the jet. This leads to density fluctuations of short wavelengths in the jet and to variability timescales of order of $ \sim 10 $ ms\footnote{While our convergence tests do not show differences in the variability timescales that are induced by the instabilities, it is possible that those are just an upper limit if the minimum wavelength of the instabilities is not resolved in the simulations.}. 
	In intermittent mildly magnetized jets the boundary instabilities are quenched due to the presence of magnetic field and the variability is governed by the timescales of the jet modulations, which are determined by the physics of the launching mechanism.
	This sets a strict constraint on the engine variability timescales in mildly magnetized GRB jets, and possibly in highly magnetized jets as well. The engine must be intermittent on timescales that are equivalent to the variability time of the prompt emission, an order of $\sim 10 $ ms. 
	Simulating such short time scales requires high grid resolutions which are beyond our computational capabilities. Thus we were limited to simulating jets with intermittent time $ \gtrsim $ 0.1 second. Our simulations indicate that jets with shorter modulations are more stable, owing to a high magnetic pressure along the jet spine that inhibits baryon entrainment from the cocoon. Thus, we expect that magnetic jets launched by engines which are intermittent on 10 ms timescales will be stable, and thus feature both the observed light curve variability and a substantial photospheric component.
	
	\begin{table*}
		\setlength{\tabcolsep}{6.8pt}
		\centering
		\begin{tabular}{ | l | c c c c | }
			
			\hline
			Jet type & Continuous hydrodynamic  & Continuous magnetized & Intermittent hydrodynamic & Intermittent magnetized \\ \hline
			Mixing source (head/boundary) & both; boundary dominates & stable & head & head \\
			Variability source (mixing/engine) & mixing & none & both & engine \\ \hline
			Variability timescales & $ \sim $ ms & \textcolor{red}{none} & set by engine and mixing & set by engine \\
			Efficiency & high & high & \textcolor{red}{$ \lesssim 1\% $} & high \\
			Spectrum broadening by shocks & \textcolor{red}{unlikely} & \textcolor{red}{no} & \textcolor{red}{no} & possibly \\

			\hline
			
		\end{tabular}
		\hfill\break
		
		\caption{
            Summary of our 3D simulation results of continuous/intermittent hydrodynamic/magnetized jets, based on this paper and on \citet{Gottlieb2019,Gottlieb2020b,Gottlieb2020a,Gottlieb2021}. Shown are the instability and variability sources of the jets, and the observables: variability, efficiency and possibility of hardening the spectrum by shocks. Red color text indicates results that are in tension with observations.
        		}
		\label{table:summary}
	\end{table*}
	
	Our analysis in this paper was limited to hydrodynamic and mildly magnetized jets and did not consider highly-magnetized jets. Numerical simulations of Poynting-flux driven jets \citep[e.g.][]{Bromberg2016,Bromberg2019} suggest that high $ \sigma $ long GRB jets dissipate their magnetic energy in narrow collimation nozzles that form close to the jet base, resulting in plasma $ \beta \sim 1 $ above the nozzles. Though kink instability should have a more significant effect on the propagation of such jets, they should feature similar stability at their boundaries to our $\sigma=0.1$ jets.
	We thus speculate that the post-breakout evolution and emission of high $ \sigma $ jets may be similar to what we find in mildly magnetized jets.

    Finally, we conclude this series of papers that explore the structure and the emission from continuous/intermittent hydrodynamic/magnetized jets by comparing our main findings (see Table \ref{table:summary}).
    Continuous hydrodynamic jets maintain ultra-relativistic $ \eta_s $ with which they generate efficient photospheric emission. Boundary instabilities lead to mixing which accounts for the observed light curve variability and triggers internal shocks. The question whether these shocks can broaden the spectrum to the observed frequencies remains unclear.
    Magnetic fields inhibit the growth of boundary instabilities and the resulting mixing. As a result, the light curve of continuous weakly magnetized jets lacks the variability that is required by observations.
    Intermittent jet launching may generate strong internal shocks that may set the hard tail of the spectrum. However, intermittent hydrodynamic jets were found to be prone to intense mixing that takes place at the head of each high-power jet episode, and reduces the radiative efficiency to be essentially zero. In this paper we showed that a magnetization of $ \sigma \gtrsim 10^{-2} $ can stabilize not only boundary instabilities, but also the head-like mixing of intermittent jets. Consequently, intermittent magnetized jets can power efficient and variable emission (set by the engine variability), that may be broaden by strong sub-photospheric internal shocks that emerge from the modulations of the engine. We therefore conclude that out of all models that we tested, an intermittent mildly magnetized jet is the most plausible one to be consistent with all observables, and thus is the leading candidate as the source of GRBs.
    
	\section*{Acknowledgements}
	
	This research is partially supported by an ERC grant (JetNS) (OG and EN).
	AL and EN acknowledge support by the Israel Science Foundation Grant No. 1114/ 17. OB was funded by an ISF grant 1657/18 and by an ISF (Icore) grant 1829/12.
	
	\section*{Data Availability}
	
	The data underlying this article will be shared on reasonable request to the corresponding author.	
	
	\bibliographystyle{mnras}
	\bibliography{main}
	
	\appendix
	\section{Convergence test} \label{sec:app}
	
	The baryon entrainment from the cocoon into the jet and the JCI is primarily taking place in the lateral direction. Therefore, our 3D RMHD simulations were chosen to maintain a very high resolution on the $ \hat{x}-\hat{y} $ plane inside the inner patch where the jet and the JCI reside. In comparison, this resolution is higher than all previous 3D simulations of magnetized jets or intermittent jets in a star \citep{LopezCamara2016,Gottlieb2020b,Gottlieb2020a}. In the parallel direction to the jet axis we keep the same cells height $ \Delta z $ of that in previous 3D simulations of magnetized jets in a star \citep{Gottlieb2020b}. The cells height is not expected to be affected by the intermittency of the jets as long as it sustains $ \Delta z \ll Tc $. This criterion is held in all of our simulations.
	
	We verify that our main conclusions are independent of the resolution by conducting a convergence test. We perform an additional simulation with an identical setup to that of $ \Wb $, since low $ \sigma $ jets do not deviate from the inner high resolution patch, such that it is easier to study the resolution effects.
	In the test simulation we double the cells resolution on the $ \hat{x} $ and $ \hat{y} $ dimensions, so that we have in total 1600 cells on each of the $ \hat{x} $ and $ \hat{y} $ axes. Since this simulation is very demanding, the $ \hat{z} $-axis stretches only up to $ R_\star $, and we compare the two simulations inside the stellar boundaries. 
	
\begin{figure}
		\centering
		\includegraphics[scale=0.21]{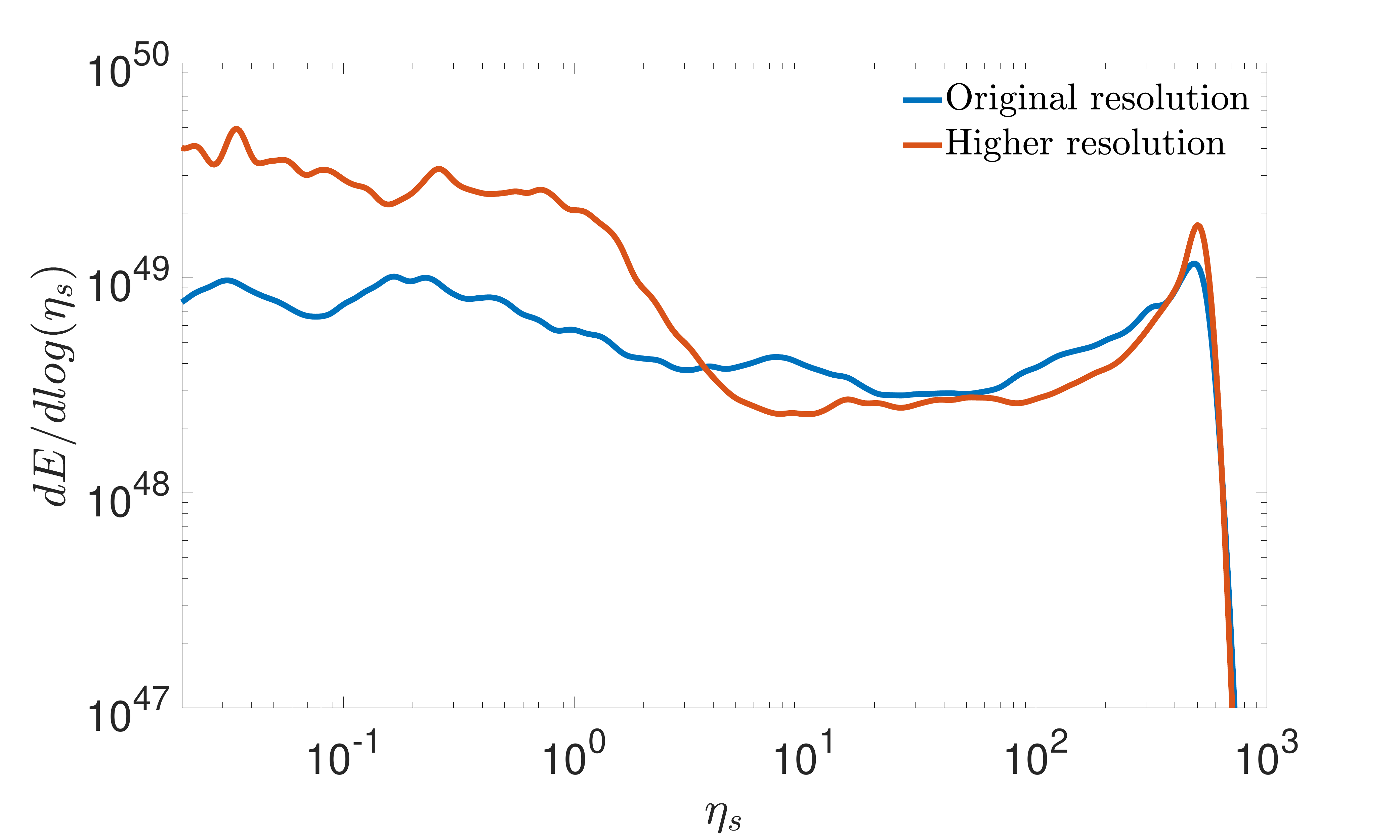}
		\caption[Convergence]{
			A comparison of the energy distribution of the original grid resolution (blue) and the higher resolution (red) as a function of $ \eta_s $. The distributions are taken when jet head reaches the stellar surface (11 s in the original simulation and 13.5 s in the simulation with the higher resolution.).
		}
		\label{fig:convergence}
	\end{figure}

	Figure \ref{fig:convergence} depicts the energy distribution per logarithmic space of $ \eta_s $ upon jet breakout from the star, indicating the jet stability inside the star. The agreement between the original simulation (blue) and the test one (red) is remarkable at $ \eta_s \gtrsim 3 $, which manifests the jet and the JCI regions \citep{Gottlieb2021} that are inside the high resolution patch. However, the similarity between the cocoons ($ \eta_s \lesssim 3 $), which reside in the outer patches in these simulations, is less compelling. The energy in the test simulation is higher since jets in higher resolutions propagate slower and thus the jet breakout is delayed in comparison to lower resolutions. Here the distribution of the test simulation is taken 13.5 s after the jet launching whereas that of the original simulation is at 11 s, and thus its total energy is also lower.
	
	In this work our focus is neither the jet breakout time nor the structure of the cocoon. Thus, the differences between the two simulations are not a concern for our purposes. The mixing in the JCI and the jet seems to be well-matched between the simulations, even though the resolutions in the lateral direction differ by a factor of two. We thus conclude that the mixing and the stability of our jets are unlikely to be subject to numerical issues.
	
	\label{lastpage}
\end{document}